\documentclass[10pt]{article}
\usepackage{graphicx}
\usepackage{subcaption}
\usepackage{float}
\usepackage{verbatim}
\usepackage{hyperref}
\usepackage{lineno}
\usepackage{amsmath}
\usepackage{amssymb}

\def\Title#1{\begin{center} {\Large #1 } \end{center}}
\def\Author#1{\begin{center}{ \sc #1} \end{center}}
\def\Address#1{\begin{center}{ \it #1} \end{center}}

\newcommand\pubblock{\rightline{\begin{tabular}{l} Proceedings of the Fifth Annual LHCP\\ \pubnumber\\
         \pubdate  \end{tabular}}}

\newenvironment{Abstract}{\begin{quotation} \begin{center} 
             \large ABSTRACT \end{center}\bigskip 
      \begin{center}\begin{large}}{\end{large}\end{center} \end{quotation}}

\newenvironment{Presented}{\begin{quotation} \begin{center} 
             PRESENTED AT\end{center}\bigskip 
      \begin{center}\begin{large}}{\end{large}\end{center} \end{quotation}}

\def\Acknowledgements{\bigskip  \bigskip \begin{center} \begin{large}
             \bf ACKNOWLEDGEMENTS \end{large}\end{center}}

\def\beq{\begin{equation}}
\def\eeq#1{\label{#1}\end{equation}}
\def\eeqn{\end{equation}}

\def\beqa{\begin{eqnarray}}
\def\eeqa#1{\label{#1}\end{eqnarray}}
\def\eeqan{\end{eqnarray}}

\let\bar=\overbar

\def\Dslash{\not{\hbox{\kern-4pt $D$}}}
\def\dslash{\not{\hbox{\kern-2pt $\del$}}}

\def\msb{{\bar{\ssstyle M \kern -1pt S}}}

\def\pt{$\it{p}_{T}$}

\usepackage{color}

\newcommand\pubnumber{ ALICE-PHYS-PROC-2017 }

\newcommand\pubdate{\today}

\def\affiliation{
On behalf of the ALICE Collaboration, \\
Instituto de Ciencias Nucleares, Universidad Nacional Aut\'onoma de M\'exico, \\
Apartado Postal 70-543, M\'exico Distrito Federal 04510, M\'exico}

\begin{document}

\large
\begin{titlepage}
\pubblock

\vfill
\Title{  New results on collectivity with ALICE  }
\vfill

\Author{ Omar V\'azquez Rueda  }
\Address{\affiliation}
\vfill
\begin{Abstract}

An overview of recent ALICE results aimed to understand collective phenomena in Pb--Pb collisions at the LHC is presented. These include the centrality dependence of the transverse momentum ($p_{\rm T}$) distributions of charged pions, kaons, and protons as well as results on anisotropic flow using data from Pb--Pb collisions at $\sqrt{s_{{\mathrm{NN}}}} = 5.02$~TeV. In addition,  comparisons between data and model predictions are presented. 

\end{Abstract}
\vfill

\begin{Presented}
The Fifth Annual Conference\\
 on Large Hadron Collider Physics \\
Shanghai Jiao Tong University, Shanghai, China\\ 
May 15-20, 2017
\end{Presented}
\vfill
\end{titlepage}
%

\normalsize 


\section{Introduction}

Ultrarelativistic heavy-ion collisions at the Large Hadron Collider (LHC) produce a strongly interacting Quark--Gluon Plasma \cite{Aad:ReviewOfLHCRun1Results, Aad:FlowandJetQuenching1}. The study of the production of charged pions, kaons, and protons from low to intermediate transverse momentum ($p_{{\rm T}} \lessapprox $ 10 GeV/$c$) provides information about collective phenomena and hadronization mechanisms like recombination \cite{Aad:Recombinationpaper}. Moreover, particle production above 10 GeV/$c$ can give information about parton energy loss mechanisms within the medium \cite{Aad:FlowandJetQuenching2, Aad:RAApaper, Aad:SlossPaper}.\\

ALICE (A Large Ion Collider Experiment) is the dedicated heavy-ion experiment at the LHC with unique capabilities for tracking and particle identification over a wide range of transverse momentum, ranging from approximately hundreds of MeV/$c$ up to tens of GeV/$c$. In this paper a brief review of selected results obtained using the Pb–Pb data sample collected by ALICE during the LHC Run 2 at the unprecedent energy of $\sqrt{s_{{\rm NN}}} = 5.02$~TeV will be presented. The results include the centrality dependence of the transverse momentum distributions of identified charged particles, which are used to quantify radial flow effects. Measurements of anisotropic flow and comparisons with hydrodynamical predictions are also presented. The paper is organized as follows. In section 2, the ALICE apparatus as well as the techniques used to perform the measurements are briefly discussed. Results and discussions are presented in section 3. Final remarks are summarized in section 4.

\section{ALICE apparatus}

The Pb--Pb data sample at $\sqrt{s_{{\rm NN}}} = 5.02$~TeV was collected with the ALICE detector during LHC Run 2 in 2015 data taking. The measurement of the $p_{{\rm T}}$ spectra was accomplished using about 9.5 million events recorded at low interaction rate. The HMPID (High Momentum Particle Identification Detector) analysis (due to its smaller acceptance) was based on a larger sample of 100 million events recorded at high interaction rate, which amount to 100 million events.\\ 
The trigger selection was accomplished using the V0 detector, which is composed of two arrays of scintillator counters (V0A and V0C detectors). The triggering corresponds to a logical AND between signal inputs from V0A and V0C detectors and it is based on the V0 timing signal. The triggered events were classified in nine different centrality classes.\\

The measurement of the $p_{{\rm T}}$ spectra  at midrapidity ($|y|<0.5$) employs different detectors of the ALICE central barrel system. Full particle identification (PID) capabilities of the ITS, TPC, TOF, and HMPID detectors have been exploited by six independent analyses. Table~\ref{tab:pTranges} lists the \pt~ intervals in which each analysis contributes to the combined pion, kaon, and proton spectra. The ITS (Inner Tracking System) is composed of six layers of silicon detectors, it is used for triggering, tracking, and vertex reconstruction. In the standalone analysis with the ITS (ITSsa) PID is performed via the specific energy loss (${\rm d}E/{\rm d}x$) whose resolution is about 6\%. The TPC (Time Projection Chamber) is the main tracking device of ALICE~\cite{Aad:ALICEexperiment}, allowing also for particle identification by means of the ${\rm d}E/{\rm d}x$. In a high particle density enviroment, the resolution of the ${\rm d}E/{\rm d}x$ is about 5\%.  In the momentum region where the ${\rm d}E/{\rm d}x$ signal among all the species is well separated, track--by--track identification is performed by parameterizing the number of sigmas distribution (defined as the relative difference between the measured $\langle {\rm d}E/{\rm d}x \rangle$ and the expected from the Bethe-Bloch curve) with a Gaussian function. The relativistic rise analysis (rTPC) starts at 2(3) GeV/$c$ for pions (kaons and protons)~\cite{Aad:RAApaper}. In the relativistic rise of the ${\rm d}E/{\rm d}x$ in the TPC, the $\langle {\rm d}E/{\rm d}x \rangle$ increases as $log(\beta \gamma)$ in the interval ( 3 $ < \beta \gamma < $ 1000 ). Moreover the nearly constant separation among pions, kaons, and protons allows us to exploit the precise knowledge of the Bethe-Bloch curve. Particles were identified by fitting with the sum of four Gaussian functions the ${\rm d}E/{\rm d}x$ distribution for each momentum interval where the mean values of the Gaussians (for pions, kaons, protons, and electrons)  were fixed based on the Bethe--Bloch curve. Particle identification with the TOF (Time Of Flight Detector) detector is based on the Time--Of--Flight measurement. The particle yields were extracted from template fits to the TOF signal. The HMPID~\cite{Aad:HMPIDexperiment} detector consists of seven identical proximity focusing RICH (Ring Imaging Cherenkov) counters. Photon and charged particle detection is provided by a Multi-Wire Proportional Chamber (MWPC) coupled to a CsI photocathode. PID was achieved by parameterizing the Cherenkov angle ($\theta_{{\rm CH}}$) distribution with the sum of three Gaussian functions. Each Gaussian function  describes the production of pions, kaons, and protons in a given momentum interval.  The kinks analysis studies the distinctive kink topology of the weak decay of charged kaons as an alternative measurement of the charged kaon transverse momentum spectra.

\begin{table}[t]
\begin{center}
\begin{tabular}{c|ccc}  
\textbf{Analysis} &   \textbf{$\pi$}  &  \textbf{K} &  \textbf{p} \\ \hline
\textbf{ITSsa}& 0.10--0.70 &  0.20--0.50 &  0.30--0.60 \\ \hline
\textbf{TOF}  & 0.60--2.50 &  1.0--2.50  &  0.80--4.0  \\ \hline
\textbf{TPC}  & 0.25--0.70 &  0.25--0.45 &  0.45--0.80 \\ \hline
\textbf{rTPC} & 2.0--12.0  &  3.0--12.0  &  3.0--12.0 \\ \hline
\textbf{kinks}&     --     &  0.20--5.0  &     -- \\ \hline
\textbf{HMPID}& 1.50--4.0  &  1.50--4.0    &  1.50--6.0 \\ \hline
\end{tabular}
\caption{Transverse momentum ranges in GeV/$c$ for pions, kaons, and protons used by the different analyses.}
\label{tab:pTranges}
\end{center}
\end{table}

\section{Results and discussions}

\subsection{Transverse momentum spectra}

\begin{figure}[H]
    \centering
    \begin{subfigure}[b]{0.28\textwidth}
        \includegraphics[width=\textwidth]{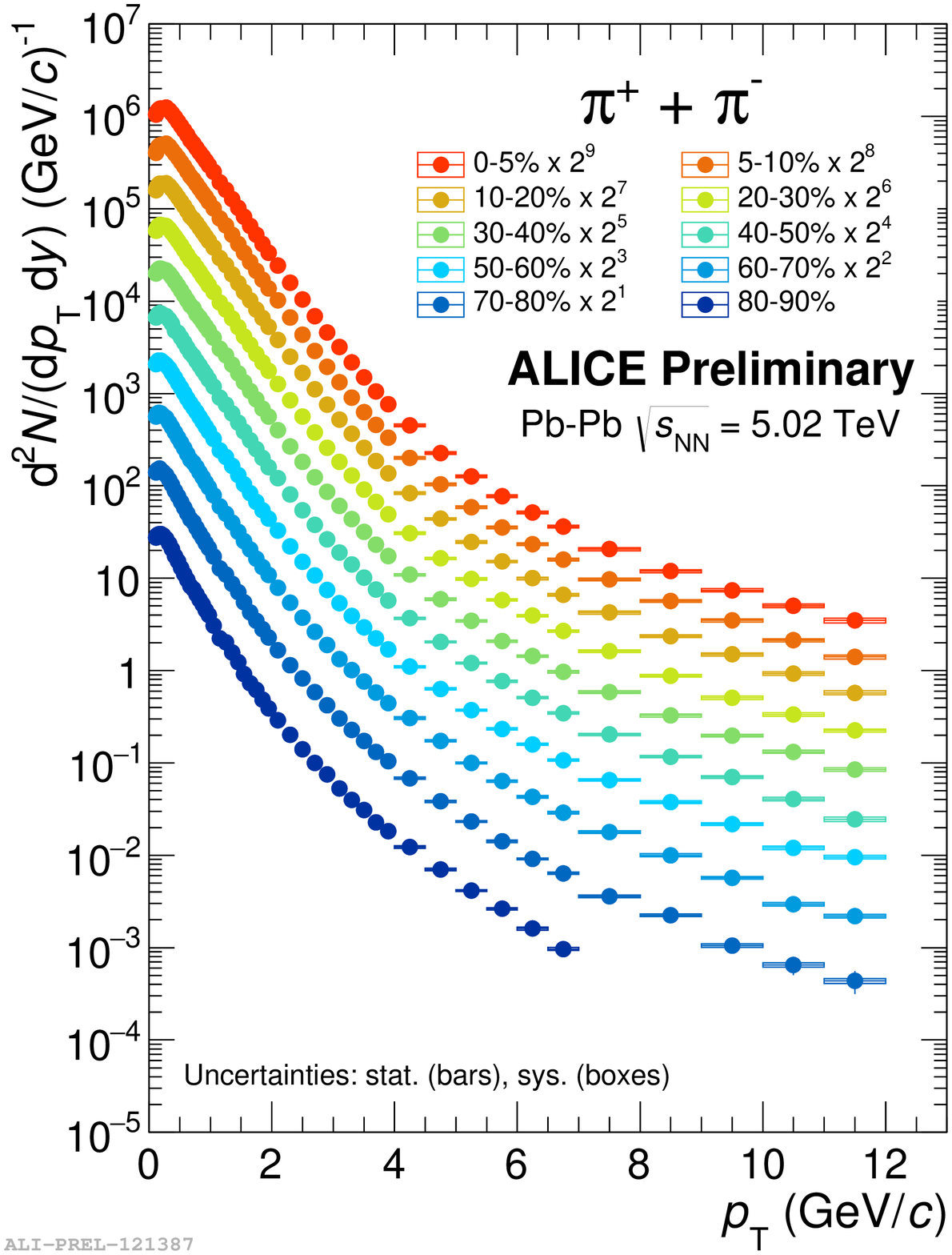}
        \label{fig:gull}
    \end{subfigure}
    \begin{subfigure}[b]{0.28\textwidth}
        \includegraphics[width=\textwidth]{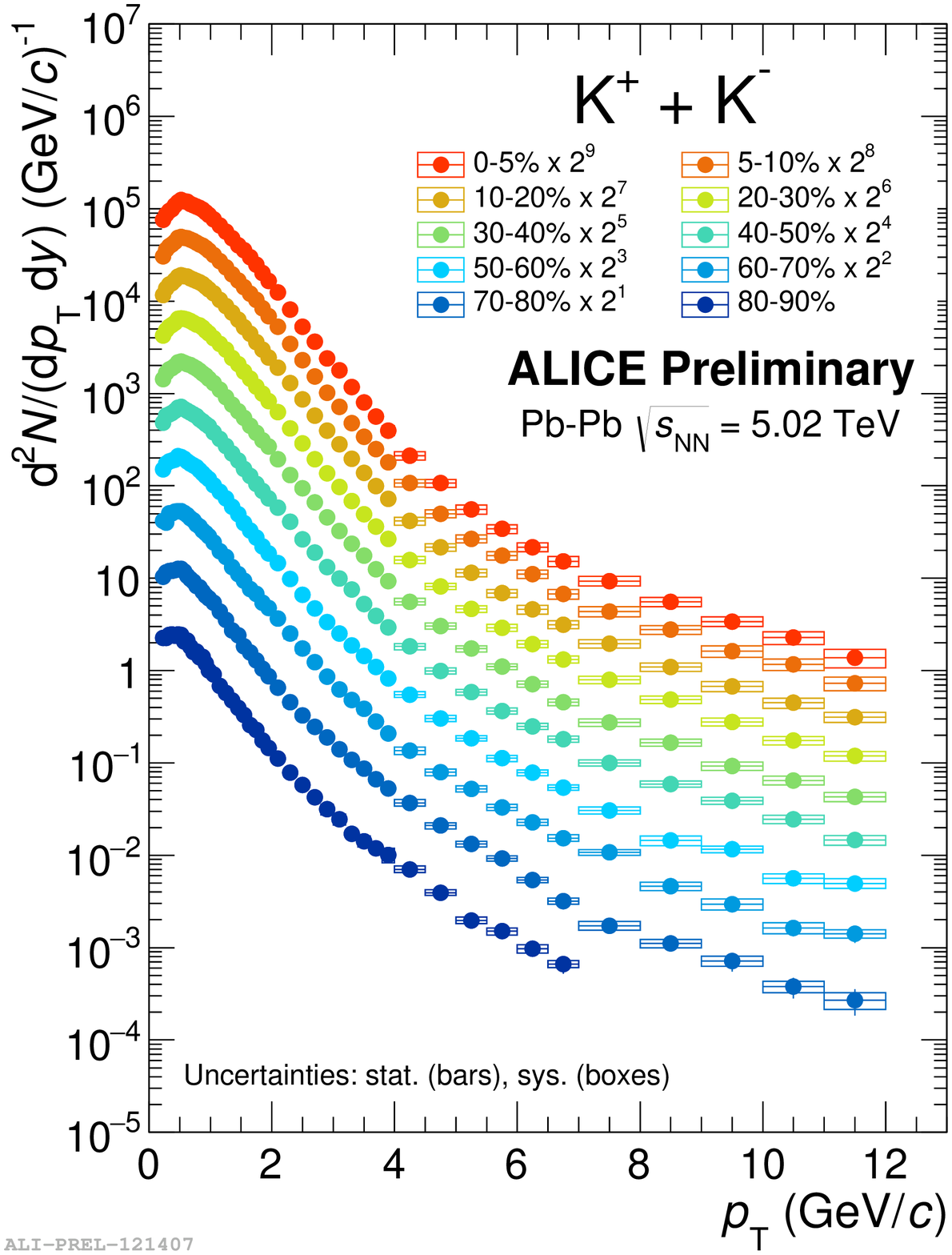}
        \label{fig:tiger}
    \end{subfigure}
    \begin{subfigure}[b]{0.28\textwidth}
        \includegraphics[width=\textwidth]{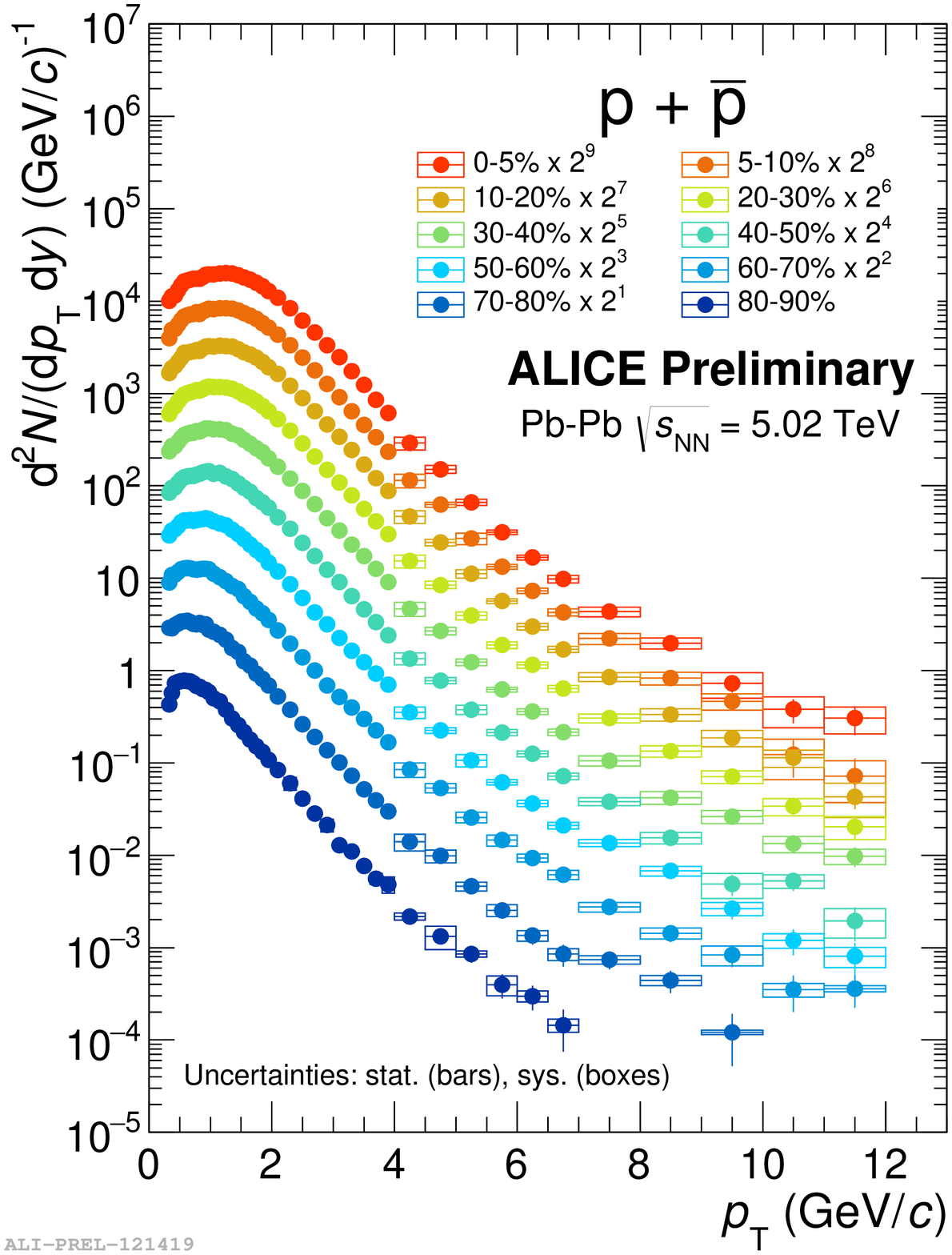}
        \label{fig:mouse}
    \end{subfigure}
    \caption{Transverse momentum spectra of pions (left), kaons (middle), and protons (right) from Pb--Pb collisions at $\sqrt{s_{{NN}}} = 5.02$~TeV for different centrality classes. Statistical and systematic uncertainties are displayed, the former are shown as bars while the latter as empty boxes.}
    \label{Fig:PIDSpectra}
\end{figure}

\begin{figure}[!h]
    \centering
         \includegraphics[width=0.71\textwidth]{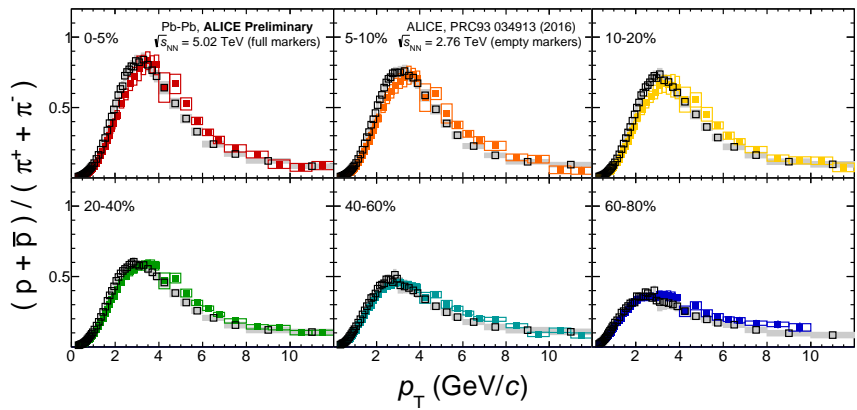}
    \caption{Proton-to-pion ratio of transverse momentum spectra for different centrality classes. The most central (peripheral)  collisions are displayed in the top-left (bottom-right) corner. Systematic uncertainties are shown as empty boxes.}
        \label{fig:ProtonToPion}
\end{figure}

Figure \ref{Fig:PIDSpectra} shows the transverse momentum spectra of charged pions, kaons, and protons for ten centrality classes. Going from peripheral to central collisions, a hardening of the spectral shapes is observed at $p_{{\rm T}}$ $<$ 2 GeV/$c$. This is a mass-dependent effect, which is more significant for protons. The observed behavior is consistent with the expected effects due to radial flow~\cite{Aad:RadialFlow}. A similar example of a behavior dominated by radial flow is visualized in Fig.~\ref{fig:ProtonToPion} where the centrality dependence of the proton-to-pion ratio as a function of transverse momentum in Pb--Pb collisions at $\sqrt{s_{{\rm NN}}} = 5.02$~TeV is shown. The ratio exhibits a maximum in the $p_{{\rm T}}$ range 2--10 GeV/$c$. The size of the effect increases going from the most peripheral to the most central Pb--Pb collisions. Moreover, a comparison with results at $\sqrt{s_{{\rm NN}}} = 2.76$~TeV shows that the position of the peak is shifted towards higher values of $p_{{\rm T}}$ due to the presence of stronger radial flow.  

\begin{figure*}[t!]
\begin{center}
    \includegraphics[width=0.27\textwidth]{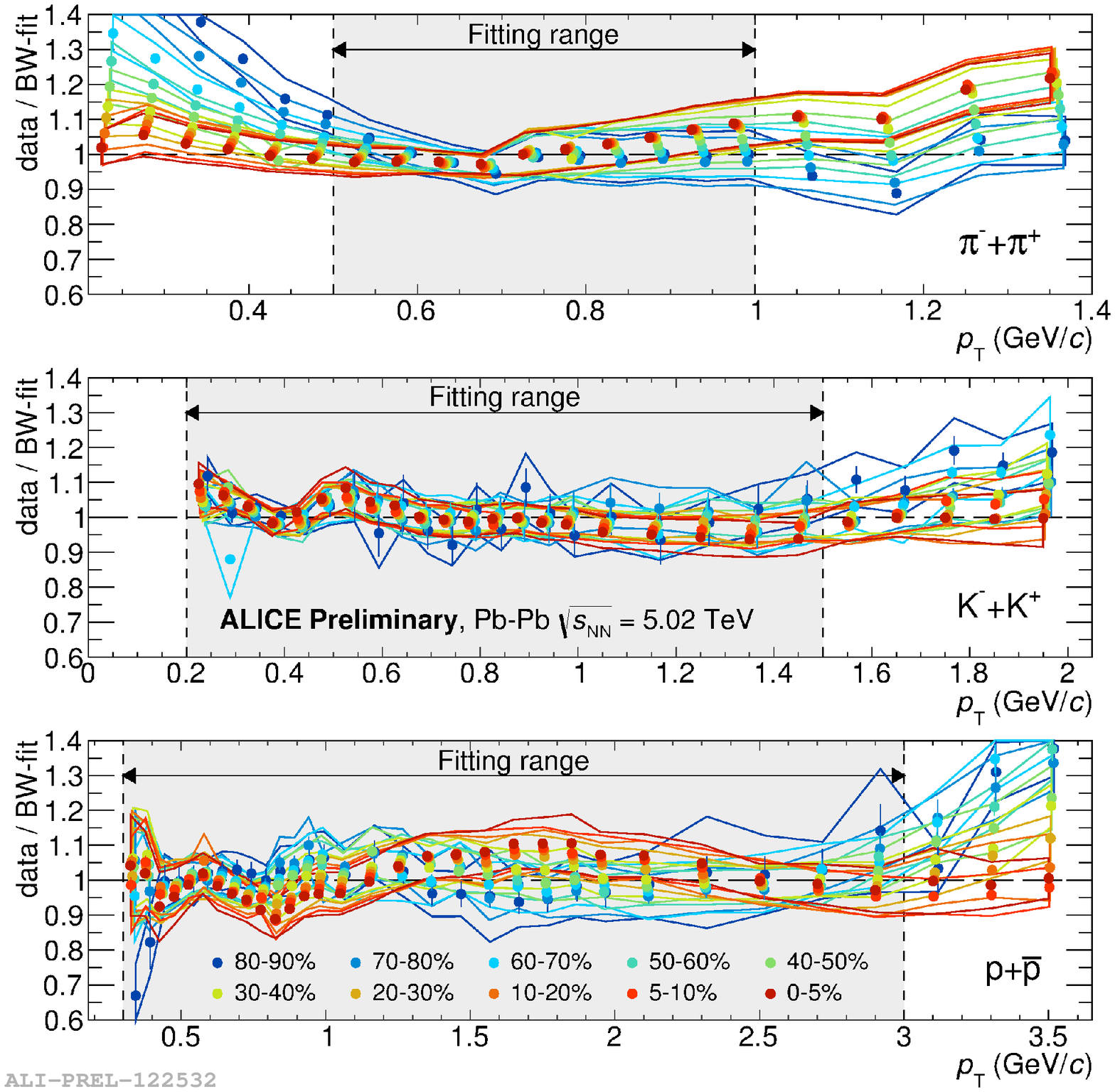}    
    \includegraphics[width=0.37\textwidth]{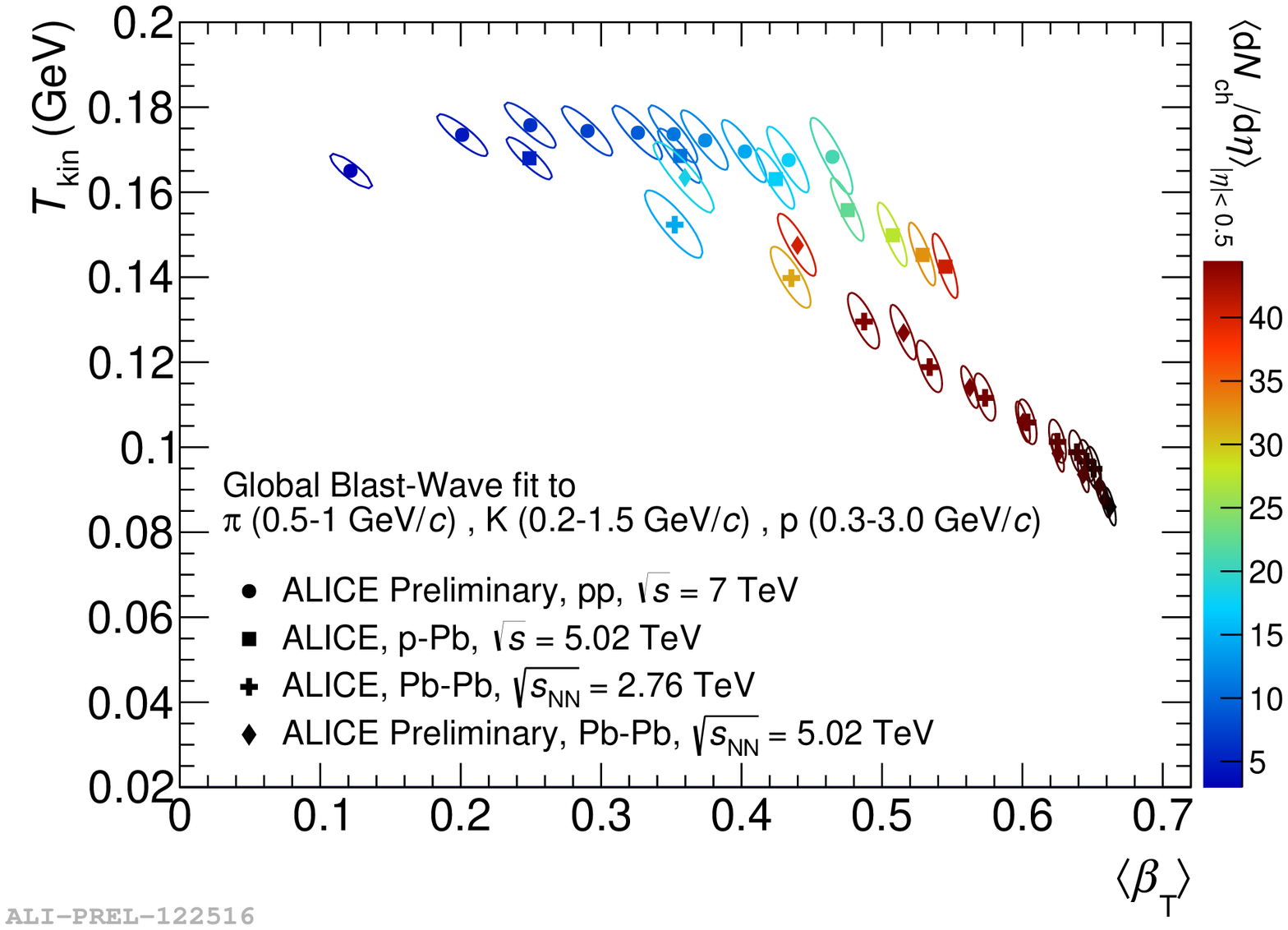}    
\caption{(Left) The transverse momentum distributions of charged pions (top), kaons (middle), and protons (bottom) normalized to the corresponding blast-wave functions, different centralities are shown. (Right) $T_{{\rm kin}}$ vs $\langle \beta_{{\rm T}} \rangle$ results of the Blast--Wave analysis, results for Pb--Pb at $\sqrt{s_{{\rm NN}}} = 5.02$~TeV are represented as rhombs for different collision centralities. The color scale represents the mean multiplicity density in $|\eta|<0.5$ for different colliding systems. } 
        \label{Fig:BWfit}
\end{center}
\end{figure*}

In order to quantity the change of the spectral shapes with collision centrality, a combined fit with a Blast-wave function has been performed. The functional form of the model~\cite{Aad:BWModel} is given by

\begin{eqnarray}
\frac{1}{p_{{\rm T}}}\frac{{\rm d}N}{{\rm d}p_{{\rm T}}}\propto \int_{0}^{{\rm R}} m_{{\rm T}}I_{0}\left(\frac{p_{{\rm T}}\sinh \rho}{T_{{\rm kin}}}\right)K_{1}\left(\frac{{ m}_{{\rm T}}\cosh\rho}{T_{{\rm kin}}}\right)rdr,
\end{eqnarray}

\noindent where $I_{0}$ and $K_{1}$ are the modified Bessel functions, $\rho=\tanh^{-1}(\beta_{T}) =\tanh^{-1}\left(\left(\frac{r}{R}\right)^{n}\beta_{s}\right)$, 
where $r$ is the radial distance from the center of the fireball in the transverse plane, $R$ is the radius of the fireball, $\beta_{s}$  is the transverse expansion velocity at the surface and $n$ is the exponent of the velocity profile. $T_{{\rm kin}}$, $\beta_{{\rm T}}$, and $n$ are free parameters of the fit. As the values of the free parameters are sensitive to the fit range for charged pions due to the large contribution from resonance decays (mostly at low $p_{{\rm T}}$), which tend to reduce $T_{{\rm kin}}$, the $p_{{\rm T}}$ ranges 0.5-1 GeV/$c$, 0.2-1.5 GeV/$c$ and 0.3-3 GeV/$c$ were used for pions, kaons, and protons, respectively. Figure ~\ref{Fig:BWfit} (left) shows the data-fit ratio in which a better agreement is seen for central events. Figure ~\ref{Fig:BWfit} (right) shows the correlation between $\langle \beta_{{\rm T}} \rangle$ and $T_{{\rm kin}}$, where it can be seen that going from peripheral to central Pb--Pb collisions the $\langle \beta_{{\rm T}} \rangle$  ($T_{{\rm kin}}$) increases (decreases). Moreover for the most central Pb--Pb collisions the radial flow ($\langle \beta_{{\rm T}} \rangle$) is found to be higher in $5.02$~TeV than in $2.76$~TeV Pb--Pb collisions. For central collisions, the temperature at the kinetic freeze–out is below the expected value for the QCD phase transition ($\approx$ 157 MeV/c)\cite{Aad:gasPhaseTransition}. Therefore, the results could give indirect evidence of the existence of a hadronic gas phase within the severe limitations of the blast-wave model.

\subsection{Anisotropic flow}
\begin{figure}[H]
    \centering
        \includegraphics[height=2.3in]{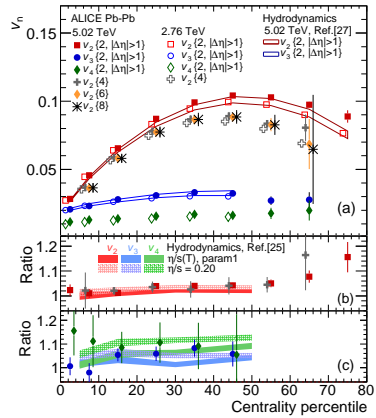}
        \caption{(a) Anisotropic flow $v_{n}$ integrated over the $p_{{\rm T}}$ range 0.2 $ < p_{{\rm T}} < $ 5.0 GeV/$c$, as a function of event centrality, for the two-particle (with $|\Delta \eta |>1$) and multi--particle correlation methods. Measurements for Pb--Pb collisions at $\sqrt{s_{{\rm NN}}} = 5.02$ and $2.76$~TeV are shown by solid and open markers, respectively. (b) Ratios of $v_{2}\{2,|\Delta \eta|>1\}$ and $v_{2}\{4\}$ results for Pb--Pb collisions at $\sqrt{s_{{\rm NN}}} = 5.02$ and $2.76$~TeV. (c) Ratios of $v_{3}\{2,|\Delta \eta|>1\}$ and $v_{4}\{2,|\Delta \eta|>1\}$ measurements for Pb--Pb collisions at $\sqrt{s_{{\rm NN}}} = 5.02$ and $2.76$~TeV. Figure taken from~\cite{Aad:Flow5Tev}.}
        \label{Fig:AnisotropicFlowChargedParticles}
\end{figure}

Anisotropic flow in heavy-ion collisions is understood as the hydrodynamic response of the produced matter to the spatial anisotropy energy density profile in the early stages of the collision. Anisotropic flow ($v_{n}$) is characterized by the coefficients of the Fourier series decomposition of the azimuthal distribution

\begin{equation}
\centering
\frac{{\rm d}N}{{\rm d}\varphi}\propto 1 + 2\sum_{n=1}^{\infty}v_{n}\cos[n(\varphi - \Psi_{n})]
\end{equation}

\noindent
where $\Psi_{n}$ are the corresponding symmetry planes~\cite{Aad:FourierDecomposition}. Measurements of anisotropic flow are biased by fluctuations and non-flow effects. To improve anisotropic flow measurements, studies on multi--particle correlations (cumulants) are carried out as proposed in~\cite{Aad:AnisotropicFlow2}. Figure \ref{Fig:AnisotropicFlowChargedParticles}(a) shows the centrality dependence of $v_{2}, v_{3}$, and $v_{4}$ for two-particle correlations and cumulants integrated over the $p_{{\rm T}}$ range $0.2 < p_{{\rm T}} < 5.0$ GeV/$c$ for Pb--Pb collisions at $\sqrt{s_{{\rm NN}}} = 5.02$ and $2.76$~TeV. In two-particle correlations, $v_{2}\{2,|\Delta \eta|> 1\}$, a gap in pseudorapidity ($|\Delta \eta|>1$) between the two particles is applied to decrease non-flow effects, but nevertheless differences of $v_{2}$ for two-particle and cumulants remain, this effect can be related to elliptic flow fluctuations \cite{Aad:AnisotropicFlowFluctuations}. In contrast, results obtained from cumulants (such as $v_{2}\{4\}, v_{2}\{6\},$ and $v_{2}\{8\}$, which are correlations for four, six, and eight particles, respectively) were found to be consistent within 1\%, which indicates that many-particle correlations suppress non-flow effects. The general behavior of elliptic flow is that it increases from central to peripheral collisions, reaching a maximum value of 0.104 $\pm$ 0.001 (stat.) $\pm$ 0.002 (syst.) in the 40--50\% centrality class. Figure \ref{Fig:AnisotropicFlowChargedParticles}(b) shows the ratio of $v_{2}\{2,|\Delta \eta|>1\}$ for two colliding energies, $2.76$ and $5.02$~TeV. Neither energy nor centrality dependence was observed. A similar trend is observed in Fig.~\ref{Fig:AnisotropicFlowChargedParticles}(c) where the same ratio for $v_{3}\{2,|\Delta \eta|>1\}$  and $v_{4}\{2,|\Delta \eta|>1\}$ is shown. In addition hydrodynamic calculations \cite{Aad:AnisotropicFlowHydroComparison}, which consider both, a constant shear viscosity ($\eta/s$ = 0.20) and a temperature dependent $\eta/s({\rm T})$ can describe the increase in anisotropic flow. However, the measurements seem to be slightly in favor of a constant $\eta/s$ from $\sqrt{s_{{\rm NN}}} = 2.76$ to $5.02$~TeV as can be seen from the ratios in Fig.~\ref{Fig:AnisotropicFlowChargedParticles}(b) and Fig.~\ref{Fig:AnisotropicFlowChargedParticles}(c). 

\subsection{Conclusions}

We have presented a set of measurements, which used the Pb--Pb data sample at $\sqrt{s_{{\rm NN}}} = 5.02$~TeV. As compared to Pb--Pb collisions at $2.76$~TeV, a stronger radial flow at $5.02$~TeV was observed. From the Blast--Wave study, an increase of approximately 3\% of the transverse expansion velocity at $5.02$~TeV with respect to $2.76$~TeV was observed. It was observed that the increase of $v_{2}$, $v_{3}$, and $v_{4}$ from $2.76$ to $5.02$~TeV Pb--Pb collisions is more consistent with the hydrodynamical approach of anisotropic flow when a constant value of $\eta/s$ is used rather than a temperature dependent $\eta/s$. 
  
\Acknowledgements
Support for this work has been received from CONACyT under the grant number 280362.

\end{document}